\documentclass[conference]{IEEEtran}
\IEEEoverridecommandlockouts

\usepackage[caption=false,font=footnotesize]{subfig}

\usepackage{cite}
\usepackage{amsmath,amssymb,amsfonts}
\usepackage{graphicx}
\usepackage{textcomp}
\usepackage{xcolor}
\usepackage{booktabs}
\usepackage{multirow}
\usepackage{url}
\usepackage{enumitem}
\usepackage{algorithm}
\usepackage{algpseudocode}
\usepackage{array}
\usepackage{pifont}

\usepackage[a4paper, total={184mm,239mm}]{geometry}

\begin{document}

\title{Probabilistic-bit Guided CDCL for SAT Solving using Ising Consensus Assumptions}

\author{\IEEEauthorblockN{Melki Bino}
\IEEEauthorblockA{\textit{Department of Electrical \& Computer Engineering} \\
\textit{University of Texas at Dallas, Richardson, USA}\\
Email: melki.bino@utdallas.edu}}

\maketitle
\thispagestyle{plain}
\pagestyle{plain}

\begin{abstract}
Boolean satisfiability (SAT) solvers are widely used in hardware verification, 
cryptanalysis, automatic test-pattern generation, and side-channel reasoning 
workflows. Modern conflict-driven clause-learning (CDCL) solvers are highly effective, 
but satisfiable instances may still require substantial conflict analysis and 
Boolean propagation before identifying productive regions of the search space. 
This paper studies a hybrid SAT-solving framework in which a probabilistic-bit (p-bit) 
Ising sampler proposes high-agreement literals that are passed to CDCL as temporary assumptions. 
The goal is not to replace CDCL, but to evaluate whether stochastic low-violation samples 
can reduce CDCL internal search effort while retaining correctness through CDCL fallback. 
On selected controlled-backbone random 3-SAT benchmarks, the hybrid method reduces median 
conflicts by $80.8$--$85.5\%$ and median propagations by $80.2$--$84.6\%$ relative to pure CDCL. 
The observed benefit is distribution-sensitive, suggesting that p-bit guidance is effective 
only for certain instance classes. We further report exploratory machine-learning gates 
that estimate when hybrid solving is likely to help. On the selected run, a random-forest gate 
retains $94.8\%$ of hybrid wins, indicating that lightweight gating may help avoid unproductive hybrid calls.
\end{abstract}


\section{Introduction}
\label{sec:intro}

Boolean satisfiability (SAT) asks whether a Boolean formula has an assignment
that satisfies all of its clauses. Although SAT is NP-complete, modern
conflict-driven clause-learning (CDCL) solvers can solve many large practical
instances and are widely used in hardware verification, bounded model checking,
equivalence checking, automatic test-pattern generation, cryptanalysis, and
hardware-security analysis \cite{HandbookSAT2021,Ganesh2021UnreasonableSAT}.
In security-oriented applications, SAT formulations appear in logic-locking
attacks, key recovery, hardware Trojan analysis \cite{jain2025trojan}, and
side-channel constraint solving \cite{shamsi2021circuit,ahmed2025improving}. In these settings,
satisfying assignments are often meaningful artifacts, such as secret keys,
counterexample traces, test patterns, or leakage-consistent internal states.

CDCL solvers search through the Boolean assignment space using decisions, unit
propagation, conflict analysis, clause learning, and non-chronological
backtracking \cite{MarquesSilva1996GRASP}. These mechanisms make CDCL highly
robust, but satisfiable instances can still require many conflicts and
propagations before the solver reaches a productive region of the search
space. This motivates auxiliary guidance mechanisms that can propose partial
assignments likely to agree with a satisfying assignment, while leaving final
correctness to the CDCL solver.

\begin{figure}[t]
    \centering
    \includegraphics[width=\linewidth]{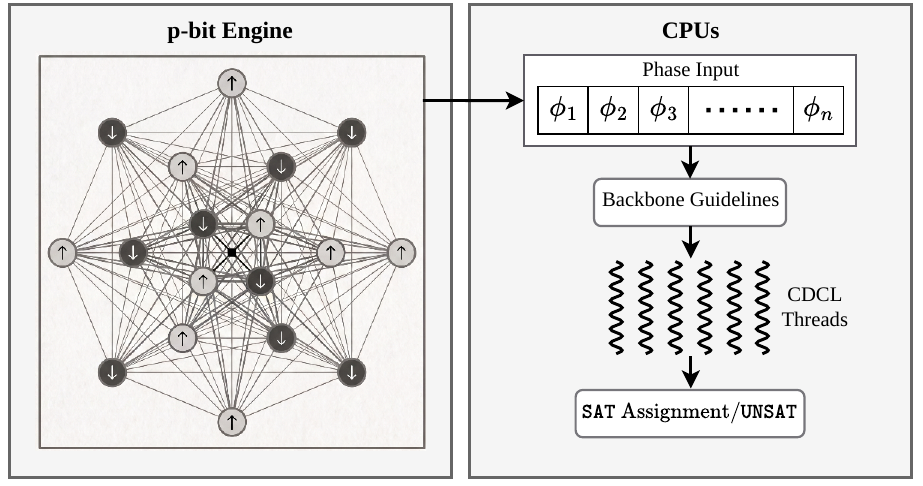}
    \caption{\textbf{Proposed p-bit-guided CDCL flow}  The p-bit Ising sampler produces
    low-violation candidate assignments whose stable consensus literals are
    passed to the CDCL solver as hard assumptions, which can reduce the initial
    CDCL subproblem.  Soundness comes from CDCL checking plus unrestricted
    rescue fallback, not from the p-bit phase.}
    \label{fig:overview}
\end{figure}

This paper studies a hybrid SAT-solving framework based on probabilistic bits
(p-bits) and Ising sampling, as shown in Fig. \ref{fig:overview}. A CNF formula
is mapped to an Ising-style energy function in which low-energy states
correspond to assignments that violate few clauses. Multiple stochastic p-bit
replicas are sampled, and high-agreement literals from low-violation samples
are passed to CDCL as temporary assumptions. CDCL then solves the reduced
subproblem under these assumptions using a bounded attempt-and-retry protocol.
The p-bit stage is heuristic and is not used as a certificate of satisfiability
or unsatisfiability. Correctness is preserved because every candidate result is
checked by CDCL, and the solver can fall back to unrestricted CDCL when the
p-bit assumptions are unproductive.

We evaluate the framework on selected satisfiable benchmark families from
SATLIB \cite{hoos2000satlib}, including random 3-SAT instances (RTI), backbone-minimal sub-instances
(BMS), and controlled-backbone random 3-SAT instances (CBS). The RTI family
contains satisfiable random 3-SAT instances with fixed numbers of variables and
clauses, while the BMS family contains corresponding backbone-minimal
sub-instances generated by removing clauses without changing the backbone. The
CBS family contains satisfiable random 3-SAT instances with controlled backbone
sizes, allowing the effect of backbone structure on solver guidance to be
studied systematically.

Our results show that p-bit-derived CDCL assumptions can substantially reduce
CDCL conflict and propagation counters on selected RTI and CBS benchmark
families. However, the benefit is distribution-sensitive. The method is less
effective on some structured or transformed instances, and BMS improvements are
partly influenced by retry and rescue-path behavior rather than by assumption
guidance alone. We therefore present the proposed method as a
study of p-bit-guided CDCL search effort, rather than as a claim of universal
SAT-solver acceleration. Our contributions are summarized as follows:

\begin{itemize}[leftmargin=10pt]
    \item We present a p-bit/Ising-assisted CDCL pipeline in which stochastic
    samples generate temporary assumptions for a conventional CDCL solver.

    \item We formulate the guidance mechanism using clause-violation energy,
    sample agreement, magnetization, and conditional CDCL search reduction.

    \item We evaluate the method on selected satisfiable SATLIB families,
    including RTI, BMS, and CBS benchmarks, and show that p-bit guidance can
    reduce CDCL conflict and propagation counters on selected RTI and CBS
    instances.

    \item We characterize distribution-sensitive behavior and identify cases
    where improvements arise from retry or rescue-path effects rather than
    direct assumption quality.

    \item We report exploratory suitability-gating results and identify feature
    leakage sources that must be removed before deployment-level evaluation.
\end{itemize}
The paper is organized as follows. Section \ref{sec:prelim} presents relevant 
preliminaries and literature context, Section \ref{sec:method} develops 
the p-bit/Ising-assisted CDCL pipeline, Section \ref{sec:exp} presents the experiments, 
and Section \ref{sec:conc} the conclusion.


\section{Preliminaries}
\label{sec:prelim}

\begin{table*}[t]
\centering
\caption{Comparison with related SAT-guidance and Ising-based solving approaches.}
\label{tab:related_work}
\setlength{\tabcolsep}{4pt}
\renewcommand{\arraystretch}{1.15}
\scriptsize
\begin{tabular}{p{2.7cm} p{4.5cm} p{1.8cm} p{3cm} c c}
\toprule
\textbf{Approach} &
\textbf{Guidance / Solver Role} &
\textbf{Platform} &
\textbf{Benchmarks} &
\textbf{Complete} &
\textbf{Gate} \\
\midrule

GNN-Core Guidance \cite{selsam2019guiding}
& UNSAT-core scores for VSIDS
& SW/GPU
& SAT Competition
& \checkmark
& -- \\

Fourier Warm Start \cite{kyrillidis2020fouriersat}
& Continuous relaxation for CDCL warm start
& CPU/GPU
& Hybrid Boolean constraints
& P
& -- \\

Electronic Ising SLS \cite{sharma2023augmenting}
& Ising-based standalone SLS
& Ising chip
& Random 3-SAT, Circuit-SAT
& --
& -- \\

GNN Backbone Hints \cite{Wang2024NeuroBack}
& Backbone-based phase hints
& SW/GPU
& SAT Competition
& \checkmark
& -- \\

FPGA p-bit Ising \cite{camsari2024all}
& Standalone p-bit Ising solver
& FPGA
& 3R3X SAT
& --
& -- \\

GPU CLS Warm Start \cite{cen2025massively}
& Parallel CLS for CDCL warm start
& GPU+CPU
& Cardinality, graph problems
& P
& -- \\

Finite-Field Guidance \cite{kim2026galoissat}
& Finite-field optimization before CDCL
& GPU+CPU
& SAT Competition 2024
& P
& -- \\

Circuit-Aware GNN \cite{cascad2025}
& Phase hints and clause filtering
& SW/GPU
& LEC circuit benchmarks
& \checkmark
& -- \\

Structured Ising SAT \cite{structuredising2025}
& Standalone Ising formulation
& Ising/SW
& Factorization SAT
& --
& -- \\


\textbf{This work}
& p-bit literals as CDCL assumptions
& SW
& Selected SATLIB
& \checkmark
& Expl. \\

\bottomrule
\end{tabular}

\vspace{0.5ex}
\begin{flushleft}
\footnotesize
\textit{Notes:} ``Complete'' indicates whether the overall method preserves SAT/UNSAT correctness through a complete CDCL procedure. 
``Gate'' indicates whether a learned suitability classifier is used to decide when guidance is applied; Expl. marks the leakage-contaminated exploratory gate in this artifact. 
P denotes partial completeness or completeness only through a CDCL fallback stage.
\end{flushleft}
\end{table*}

\textbf{CDCL SAT Solving.}
Boolean satisfiability (SAT) asks whether a Boolean formula admits an
assignment that satisfies all clauses. Modern SAT solvers are largely based on
conflict-driven clause learning (CDCL), which extends the DPLL procedure with
conflict analysis, learned clauses, and non-chronological backtracking
\cite{davis1962machine,MarquesSilva1996GRASP}. A CDCL solver maintains a
partial assignment over Boolean variables. At each step, the solver makes
branching decisions, applies Boolean constraint propagation (BCP), detects
conflicts when a clause becomes falsified, analyzes the implication graph, and
learns a new clause that prevents the same conflicting assignment from being
revisited.

Two useful indicators of CDCL search effort are the number of conflicts and the
number of propagated literals. We denote the conflict count by
\begin{equation}
    N_{\mathrm{conf}} =
    \sum_t \mathbf{1}\{\text{iteration }t\text{ derives a conflict}\},
\end{equation}
and the propagation count by
\begin{equation}
    N_{\mathrm{prop}} =
    \sum_t \#\{\text{literals assigned by BCP at iteration }t\}.
\end{equation}
These expressions are conceptual summaries of solver activity. In the
experiments, we use the accumulated solver-reported counters obtained through
CaDiCaL/PySAT, including counters accumulated during retry and rescue phases
when those phases are invoked.

Practical CDCL performance depends not only on the high-level search procedure
but also on several implementation and heuristic choices. MiniSat popularized
efficient watched-literal propagation, activity-based branching, and phase
saving \cite{Een2003ExtensibleSAT}. Glucose introduced learned-clause quality
metrics based on literal block distance (LBD) \cite{AudemardSimonIJCAI09}.
CaDiCaL incorporates modern preprocessing, inprocessing, simplification, and
restart strategies \cite{BiereFallerFazekasFleuryFroleyksPollitt-CAV24,cadical195}.
These techniques make CDCL solvers highly robust, but satisfiable instances may
still require many conflicts and propagations before a useful region of the
search space is reached. This motivates external guidance mechanisms that can
provide promising partial assignments while preserving CDCL as the final
correctness-preserving engine.

\textbf{Assumption-Based Solving.}
Most modern SAT solvers support solving under assumptions. Given a CNF formula
$F$ and a set of literals $A=\{a_1,\ldots,a_k\}$, the solver checks the
satisfiability of
\begin{equation}
    F \wedge a_1 \wedge \cdots \wedge a_k .
\end{equation}
The assumptions are temporary: they restrict the current solver call but are not
permanently added as clauses to the formula. If the formula is satisfiable under
the assumptions, the returned assignment also satisfies the original formula
$F$. If the formula is unsatisfiable under the assumptions, this does not imply
that $F$ itself is unsatisfiable; it only shows that no satisfying assignment
exists within the subspace selected by $A$. Therefore, assumption-based guidance
must include a fallback or retry mechanism when the assumptions are incorrect or
too restrictive.

This work uses the assumption interface as the connection between the p-bit
sampler and CDCL. The p-bit stage proposes literals that appear stable across
low-violation samples, and CDCL tests those literals as temporary assumptions.
Thus, the stochastic sampler guides the search subspace, but it does not replace
conflict analysis, clause learning, or satisfiability checking.

\textbf{p-bits and Ising-Style Sampling.}
A probabilistic bit, or p-bit, is a binary stochastic unit that fluctuates
between two states according to a tunable bias. Networks of interacting p-bits
can be used to sample from Ising-like energy landscapes. In an Ising
formulation, binary variables are represented as spins
$s_i \in \{-1,+1\}$, and candidate assignments are assigned an energy. Lower
energy typically corresponds to better agreement with the encoded constraints.

For SAT, a CNF formula can be mapped to an energy function in which violated
clauses contribute penalties. A simple conceptual form is
\begin{equation}
    E(s) = \sum_{C_j \in F} \mathbf{1}\{C_j \text{ is violated by } s\},
\end{equation}
where each spin assignment $s$ corresponds to a Boolean assignment. Assignments
with smaller $E(s)$ violate fewer clauses and are therefore closer to satisfying
the formula. The p-bit sampler repeatedly explores this energy landscape and
produces a collection of candidate assignments. Literals that take the same
value across many low-violation samples are treated as high-agreement literals
and are passed to CDCL as temporary assumptions.

This approach is heuristic. Low energy does not guarantee satisfiability, and
high agreement does not guarantee that a literal belongs to a satisfying
assignment. However, when the Ising landscape correlates well with the
satisfying region, these literals can reduce the CDCL search space and lower
the number of conflicts and propagations. When the landscape is misleading,
the assumptions may hurt performance; for this reason, our framework includes
retry and unrestricted fallback.

\textbf{Parallel and Learning-Augmented SAT.}
Prior work has explored several ways to improve SAT solving beyond a single
sequential CDCL search. Parallel SAT solvers use portfolio strategies, where
multiple solvers run with different heuristics, or divide-and-conquer
strategies, where the search space is partitioned into subproblems
\cite{LeFrioux2019DivideAndConquer}. GPU-based approaches
have attempted to accelerate propagation, local search, or simplification by
using parallel hardware \cite{osama2021sat,Fujii2012GPUBCP}.

Machine-learning-guided SAT methods use learned models to predict useful
solver information. Examples include satisfiability prediction, variable
branching guidance, phase selection, backbone prediction, unsatisfiable-core
prediction, and learned clause management
\cite{Selsam2019NeuroSAT,Selsam2019NeuroCore,Wang2024NeuroBack}. Other methods
use continuous or differentiable relaxations to produce warm starts or candidate
assignments for SAT solving
\cite{kyrillidis2020fouriersat,Cen2023FastFourierSAT,Zhang2024DiffSAT}.
These methods differ in how strongly they modify the underlying solver. Some
change branching or phase heuristics, some warm-start CDCL, and some operate as
standalone incomplete search procedures.

The proposed framework is closest in spirit to methods that generate auxiliary
guidance for CDCL. However, our guidance is obtained from p-bit/Ising samples
and is applied through the solver's assumption interface. CDCL remains solely
responsible for validating assignments and proving unsatisfiability. This
distinction is important because the p-bit phase is not complete and does not
provide a proof certificate by itself.

\textbf{Security-Motivated SAT.}
SAT is widely used in cryptography and hardware security because bit-level
systems can often be encoded naturally as CNF formulas. In cryptanalysis,
satisfying assignments may correspond to secret keys, preimages, or internal
cipher states \cite{soos2009extending}. In logic-locking attacks, SAT solvers
search for discriminating input patterns and key assignments that distinguish
candidate locked circuits \cite{subramanyan2015evaluating}. In automatic
test-pattern generation and Trojan analysis, SAT constraints can encode
activation conditions, propagation paths, and structural circuit properties.
Side-channel reasoning can also be formulated as a constraint-solving problem
when observed leakage restricts the set of possible internal states
\cite{kocher1999differential,mangard2007power}.

These applications motivate hybrid witness-search methods that reduce the
number of conflicts and propagations needed to find satisfying assignments.
However, security-derived SAT instances may be structurally different from
random SAT benchmarks. Cryptographic formulas often contain XOR-like parity
relations, strong diffusion, equivalent keys, and many symmetric or
near-symmetric assignments. Side-channel constraints may be noisy,
probabilistic (often modeled as random circuit learning \cite{ahmed2026oracle}), or trace-ranked rather than clean hard clauses. Such properties
can weaken the usefulness of high-agreement literals: a sampler may fail to
produce stable literal values, or the stable values may not correspond to
globally useful CDCL assumptions. Therefore, the present work should be viewed
as a step toward p-bit-guided SAT solving, with security-domain validation left
as future work.

\textbf{Comparison with Related Work.}
Table \ref{tab:related_work} compares the proposed method with representative
Ising-based, learning-guided, and hybrid SAT-solving approaches. Existing work
has explored GNN-based branching or backbone guidance, continuous relaxations
for warm starting CDCL, electronic or FPGA Ising machines, GPU-based continuous
local search, and circuit-aware learned phase guidance. These methods differ in
three important dimensions: whether the guidance is used inside CDCL or as a
standalone solver, whether SAT/UNSAT correctness is preserved by a complete
CDCL fallback, and whether the method learns when guidance should be applied.

\begin{figure*}[t]
    \centering
    \includegraphics[width=0.8\linewidth,trim={0 0.35cm 0 2.7cm},clip]{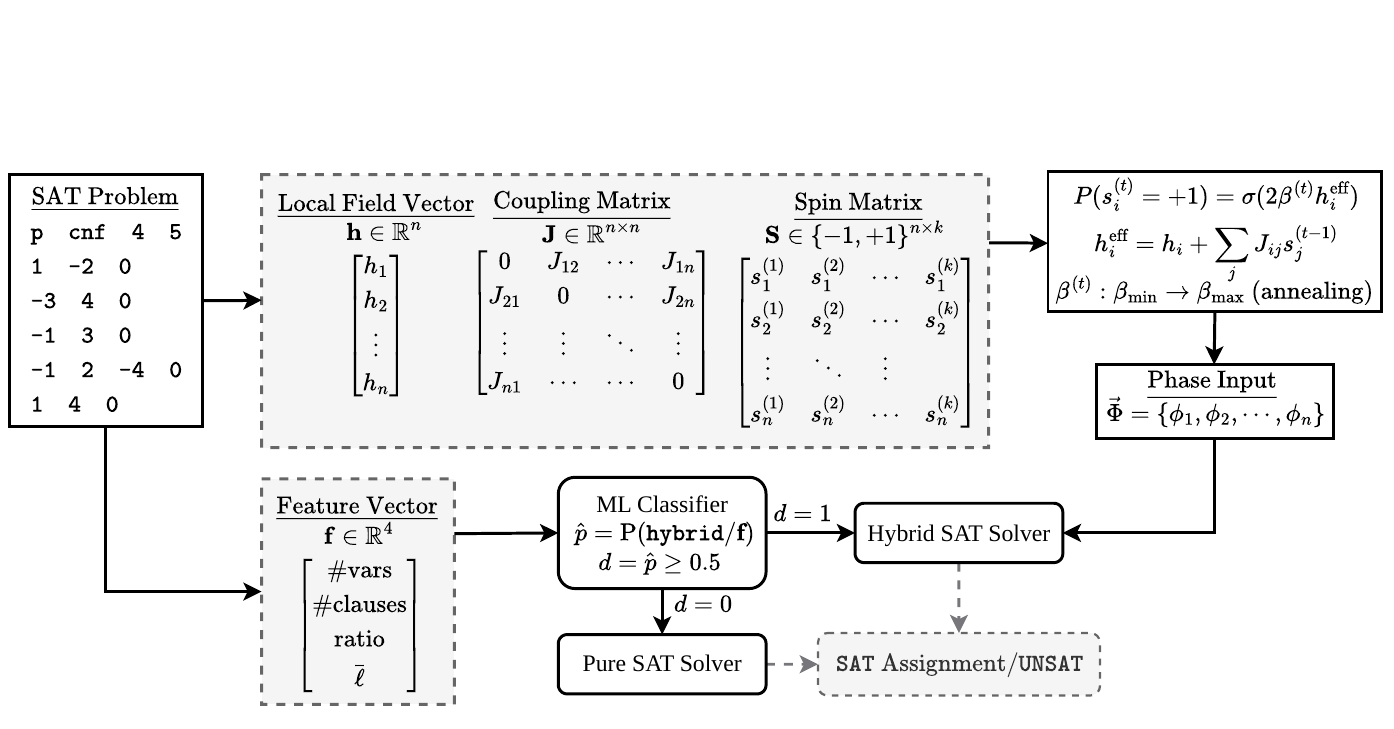}
    \caption{\textbf{Hybrid CDCL flow.}
    The input CNF is encoded as a quadratic Ising Hamiltonian. The p-bit
    engine anneals $R$ stochastic replicas from high to low temperature, and
    the resulting samples are ranked by their direct CNF violation counts.
    High-agreement literals from the best-ranked samples are passed to CDCL as
    temporary assumptions. A budget-limited attempt-and-retry stage is followed
    by unrestricted CDCL fallback, preserving completeness. An optional
    suitability gate, trained offline, determines whether a formula should use
    the hybrid path.}
    \label{fig:flow}
\end{figure*}

The present work occupies a specific point in this design space. It uses
p-bit-derived literals as temporary CDCL assumptions, retains CDCL as the
correctness-preserving backend, and studies an exploratory suitability gate for
deciding when hybrid guidance should be invoked. Unlike hardware-accelerated or
GPU-optimized approaches, our current implementation is a Python prototype and
is evaluated using CDCL internal counters rather than wall-clock speedup. The
goal of this study is therefore to evaluate whether p-bit-derived assumptions
can reduce CDCL search effort, not to claim end-to-end runtime acceleration.


\section{Probabilistic-bit Ising Guidance}
\label{sec:method}

\subsection{From CNF to Ising Energy}

Let the input CNF formula be
\begin{equation}
    F(x)=\bigwedge_{a=1}^{m} C_a(x),
\end{equation}
where $x_i\in\{0,1\}$ and each clause $C_a$ is a disjunction of literals. We
map Boolean variables to spin variables $s_i\in\{-1,+1\}$ using
\begin{equation}
    s_i = 2x_i - 1.
\end{equation}
For each clause, we define a violation indicator
\begin{equation}
    v_a(s)=
    \begin{cases}
    1, & C_a(s)\text{ is falsified},\\
    0, & C_a(s)\text{ is satisfied}.
    \end{cases}
\end{equation}
The total violation count is then
\begin{equation}
    V(s)=\sum_{a=1}^{m} v_a(s),
\end{equation}
where $V(s)=0$ if and only if $s$ satisfies the original CNF formula.

The implementation constructs a quadratic Ising Hamiltonian of the form
\begin{equation}
    E(y)=E_0+h^\top y+\frac{1}{2}y^\top Jy,
    \qquad y=[s;\,z],
\end{equation}
where $z$ denotes auxiliary spins introduced by the Rosenberg quadratization of
higher-order clause penalties. The resulting Ising model is used as a
stochastic search landscape rather than as an exact Gibbs sampler. Candidate
assignments are evaluated using the direct violation count $V(s)$ computed on
the original DIMACS clauses and original variables. Thus, auxiliary spins
influence the annealing trajectory, but they are not counted when ranking
samples for CDCL guidance.

\begin{table*}[tbh]
\setlength{\tabcolsep}{3.5pt}
\centering
\caption{SATLIB selected benchmark improvements.}
\label{tab:satlib_selected_improvements}
\begin{tabular}{|l|r|r|r|rrr|rrr|}
\hline
\multirow{2}{*}{\textbf{Benchmark}} & \multirow{2}{*}{\textbf{\#Vars}} & \multirow{2}{*}{\textbf{\#Clauses}} & \multirow{2}{*}{\textbf{Ratio}} & \multicolumn{3}{c|}{\textbf{Conflicts}} & \multicolumn{3}{c|}{\textbf{Propagations}} \\ \cline{5-10}
 &  &  &  & \textbf{Pure} & \textbf{Hybrid} & \textbf{Improve} & \textbf{Pure} & \textbf{Hybrid} & \textbf{Improve} \\ \hline
\multirow{8}{*}{CBS} & 100 & 403 & 4.03 & 286.5 & 55.0 & 80.8\% & 6693.5 & 1289.5 & 80.7\% \\ \cline{2-10}
  & 100 & 411 & 4.11 & 277.5 & 51.0 & 81.6\% & 6352.0 & 1236.5 & 80.5\% \\ \cline{2-10}
  & 100 & 418 & 4.18 & 272.5 & 47.0 & 82.8\% & 6194.5 & 1087.0 & 82.5\% \\ \cline{2-10}
  & 100 & 423 & 4.23 & 247.5 & 45.5 & 81.6\% & 5559.5 & 1103.0 & 80.2\% \\ \cline{2-10}
  & 100 & 429 & 4.29 & 245.5 & 44.0 & 82.1\% & 5600.0 & 1063.5 & 81.0\% \\ \cline{2-10}
  & 100 & 435 & 4.35 & 226.5 & 37.0 & 83.7\% & 5317.0 & 913.5 & 82.8\% \\ \cline{2-10}
  & 100 & 441 & 4.41 & 235.0 & 34.0 & 85.5\% & 5370.5 & 825.5 & 84.6\% \\ \cline{2-10}
  & 100 & 449 & 4.49 & 214.0 & 32.0 & 85.0\% & 4943.5 & 802.5 & 83.8\% \\ \hline
RTI & 100 & 429 & 4.29 & 259.5 & 43.0 & 83.4\% & 6012.0 & 1015.5 & 83.1\% \\ \hline
BMS & 100 & 429 & 4.29 & 650.5 & 404.5 & 37.8\% & 15742.5 & 9245.5 & 41.3\% \\ \hline
\end{tabular}
\end{table*}

\subsection{p-bit Sampling and Assumption Selection}

\textbf{p-bit update rule.}
A p-bit is a tunable stochastic binary unit. During sampling, each spin is
updated according to its local field. With the sign convention used by the
implementation, the effective local field is
\begin{equation}
    \tilde{\ell}_i=-h_i-\sum_j J_{ij}y_j ,
\end{equation}
and the update probability is
\begin{equation}
    \Pr(y_i=+1\mid y_{\setminus i})
    =
    \frac{1}{2}\left(1+\tanh(\beta\tilde{\ell}_i)\right).
\end{equation}
The inverse temperature $\beta$ is annealed from a hot exploratory setting to a
cold exploitative setting, and $R$ independent replicas are sampled. The
replicas are ranked by $V(s)$ instead of raw Ising energy because
quadratization penalties can affect auxiliary variables in ways that are not
directly relevant to the original CNF clauses.

\textbf{High-agreement literals.}
Let $\mathcal{T}$ denote the top-$k$ replicas with the smallest direct CNF
violation counts. For each original variable, we compute the top-sample
agreement score
\begin{equation}
    \bar{s}_i^{(k)}=\frac{1}{k}\sum_{r\in\mathcal{T}} s_i^{(r)}.
\end{equation}
Variable $i$ is selected as a high-agreement candidate when
$|\bar{s}_i^{(k)}|=1$, meaning that all top-$k$ samples assign the same value
to that variable. We avoid referring to these literals as a backbone, because
backbone information is a benchmark property and is not available to the
sampler.

The selected candidates are further ranked using a quality-weighted
magnetization score,
\begin{equation}
    m_i=\sum_{r=1}^{R} w_r s_i^{(r)},\qquad
    w_r=\frac{(1+V(s^{(r)}))^{-1}}
    {\sum_q (1+V(s^{(q)}))^{-1}}.
\end{equation}
This weighting gives a larger influence to samples that violate fewer original
CNF clauses. The top-$H$ ranked candidates are converted back to SAT literals
and form the assumption set
\begin{equation}
    \rho_{\mathrm{pbit}}=\{\ell_1,\ldots,\ell_H\}.
\end{equation}
This set is then passed to the CDCL solver through its assumption interface.

\textbf{Attempt, retry, and fallback.}
The hybrid solver first invokes CDCL under the p-bit-derived assumption set
$\rho_{\mathrm{pbit}}$ with a conflict budget $B_1$. If this attempt exhausts
its budget or the assumptions lead to an unsatisfiable restricted subproblem,
the solver retries with a second budget $B_2$. If both guided attempts fail,
the framework switches to unrestricted CDCL without assumptions. This final
fallback stage preserves completeness because the original formula is solved by
a complete CDCL solver without any p-bit restrictions.

The retry and fallback stages can also influence the measured solver counters.
In particular, failed guided attempts may add learned clauses that are reused by
the unrestricted fallback. Therefore, an observed improvement can come from two
different mechanisms: direct pruning by useful p-bit-derived assumptions, or
indirect clause reuse during rescue. The experiments distinguish these effects
when interpreting benchmark-dependent behavior.

\textbf{Why p-bit-Derived Assumptions Can Help.}
Let $\mathcal{S}$ denote the set of satisfying assignments of the CNF formula.
If the p-bit-derived assumption set $\rho_{\mathrm{pbit}}$ agrees with at least
one satisfying assignment $x^*\in\mathcal{S}$, then CDCL searches a restricted
subcube,
\begin{equation}
    \Omega(\rho_{\mathrm{pbit}})
    =
    \{x\in\{0,1\}^n : x\models \rho_{\mathrm{pbit}}\},
\end{equation}
whose size is $2^{n-|\rho_{\mathrm{pbit}}|}$ before propagation. Boolean
constraint propagation can further extend the assumption set by implying
additional assignments, producing a smaller effective search region. When the
assumptions are consistent with a useful solution basin and trigger productive
propagation chains, CDCL may explore fewer inconsistent branches. This can
reduce both the conflict count $N_{\mathrm{conf}}$ and the propagation count
$N_{\mathrm{prop}}$.

This reduction is not guaranteed. If the p-bit-derived assumptions are
inconsistent with all satisfying assignments, or if they guide CDCL into a
difficult region of the search space, the solver may spend additional effort
before recovering. The attempt, retry, and fallback protocol is therefore
necessary: it allows the framework to exploit helpful assumptions while still
recovering from misleading ones.

\subsection{Hybrid Suitability Gate}

\textbf{Label definition.}
Because p-bit guidance is distribution-sensitive, the framework also studies a
suitability gate that predicts whether the hybrid path should be used. For each
formula, we define a strict binary label
\begin{equation}
    y=\mathbf{1}\!\left[
    1-\frac{C_h}{C_p}\geq 0.20
    \;\wedge\;
    1-\frac{P_h}{P_p}\geq 0.20
    \right],
\end{equation}
where $C_p$ and $P_p$ are the pure-CDCL conflict and propagation counts, and
$C_h$ and $P_h$ are the corresponding hybrid medians over random seeds. A
formula is labeled as suitable only if the hybrid method reduces both conflicts
and propagations by at least $20\%$.

\textbf{Gate architecture.}
The intended deployment architecture is a pre-solve classifier that uses only
features computable from the CNF formula before solving. We also consider a
probe-based variant that augments structural CNF features with statistics from
a short, low-cost p-bit sampling run. In this artifact, the probe consists of
$10$ samples and $200$ sweeps.

The reported gate results should be interpreted as diagnostic rather than
deployment-ready. The models are trained using a 70/30 formula-level
in-distribution split over selected SATLIB configurations, and the feature set
contains leakage sources identified in the experimental analysis. Therefore,
the current gate results are best viewed as an upper-bound indication that
hybrid suitability may be learnable, not as a validated classifier for unseen
benchmark families.

\textbf{Agreement statistic.}
The main probe statistic is the mean absolute weighted magnetization,
\begin{equation}
    q_{\mathrm{abs}}=\frac{1}{n}\sum_{i=1}^{n}|m_i|.
\end{equation}
This statistic summarizes how strongly the p-bit samples polarize across
variables. In this paper we denote it by $q_{\mathrm{abs}}$ to avoid 
confusion with the squared Edwards--Anderson order parameter.

A large $q_{\mathrm{abs}}$ indicates strong sample agreement, but it is not
sufficient to guarantee useful CDCL assumptions. For example, some instance
families can produce highly polarized p-bit samples while still yielding poor
hybrid performance. This shows that agreement strength and guidance quality are
related but distinct quantities, motivating the use of a learned suitability
gate rather than a fixed agreement threshold.

\begin{figure*}[t]
    \centering
    \subfloat[]{\includegraphics[width=0.49\linewidth]{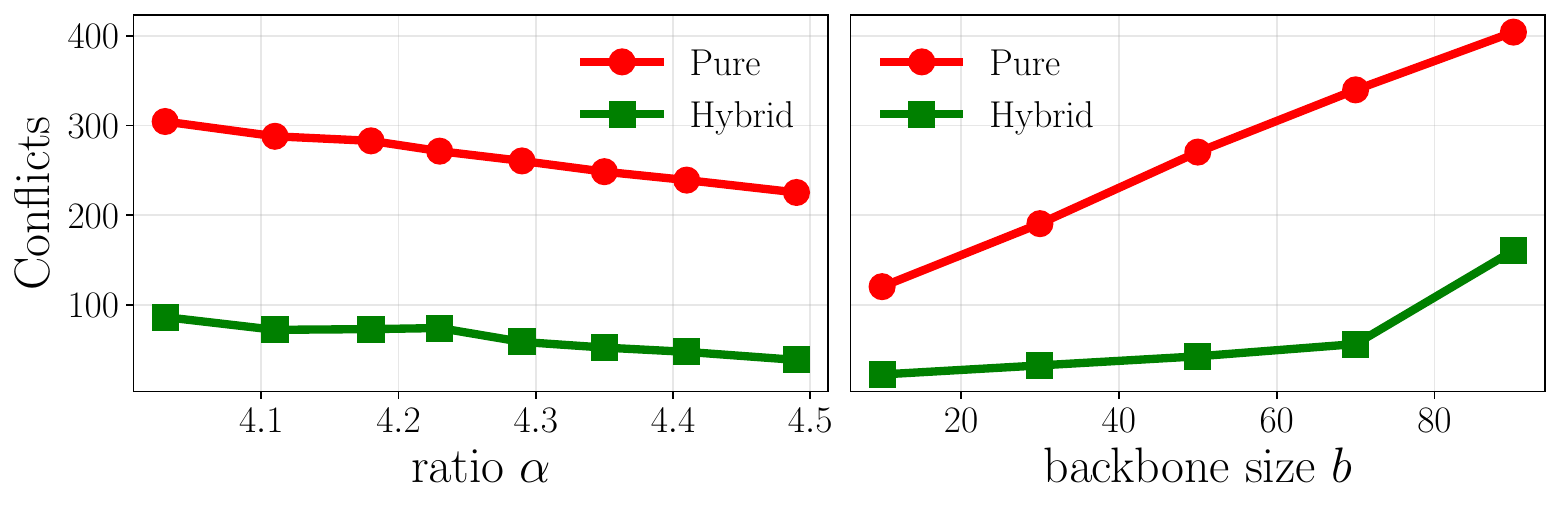}%
        \label{fig:cbs_conflicts}}%
    \hfill
    \subfloat[]{\includegraphics[width=0.49\linewidth]{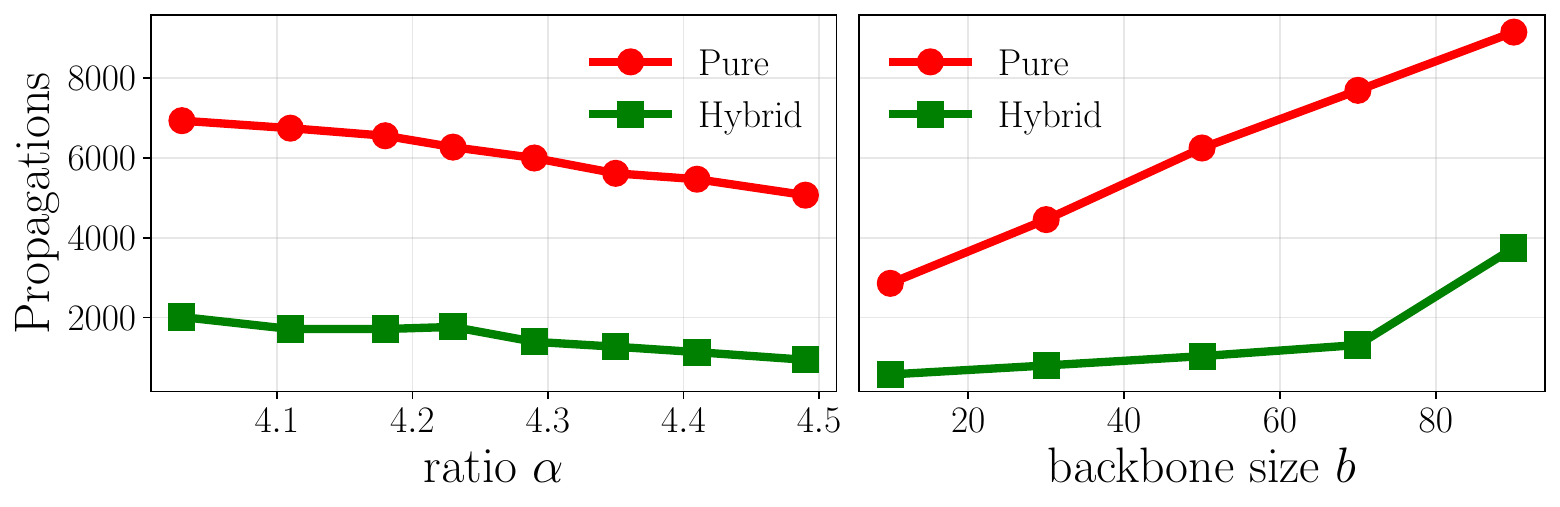}%
        \label{fig:cbs_props}}
    \caption{\textbf{Conflict and propagation reduction trends.}
    Pure and hybrid CDCL counters are averaged over CBS configurations as a
    function of clause density $\alpha=m/n$ and controlled backbone size $b$.
    All CBS instances have $n=100$ variables. The hybrid method consistently
    reduces both median conflicts and median propagations across the evaluated
    parameter sweep.}
    \label{fig:cbs_trends}
\end{figure*}

\section{Experiments}
\label{sec:exp}

\subsection{Experimental Setup}

We evaluate the proposed p-bit-guided CDCL framework on selected satisfiable
benchmark families from SATLIB \cite{hoos2000satlib}. The benchmark set contains
4,800 CNF formulas drawn from five families: controlled-backbone random 3-SAT
(CBS), random 3-SAT instances (RTI), backbone-minimal sub-instances (BMS), flat
graph-coloring instances (flat), and small-world graph-coloring instances (sw).
The CBS subset contains 4,000 formulas across 40 parameter configurations, with
all CBS instances having $n=100$ variables. The RTI and BMS subsets each contain
100 formulas, while the flat and sw graph-coloring subsets each contain 300
formulas.

The experiments were performed on a machine with
an Intel Core Ultra 7 258V processor. The processor has 8 physical cores, one
hardware thread per core, a maximum clock frequency of 4.8 GHz, 14 MiB aggregate
L2 cache, and 12 MiB L3 cache. All experiments were run on an x86\_64 Linux
environment. The SAT backend is CaDiCaL accessed through PySAT.

Each formula is solved once using pure CDCL and five times using the hybrid
p-bit-guided CDCL configuration with different random seeds. The hybrid
configuration uses $R=30$ p-bit samples, 700 annealing sweeps, top-$5$
sample agreement, at most $H=12$ CDCL assumptions, an initial guided conflict
budget of $B_1=1500$, a retry budget of $B_2=1000$, and unrestricted CDCL
fallback. Unless otherwise stated, hybrid results are reported as medians over
the five random seeds.

The primary metrics are the accumulated CaDiCaL conflict and propagation
counters. These counters include work performed during guided attempts, retries,
and unrestricted fallback when fallback is invoked. We focus on solver-internal
counters rather than wall-clock time because the current p-bit sampler is a
Python prototype and its implementation overhead would obscure the algorithmic
effect of the proposed guidance mechanism.

For the suitability-gate experiments, we use a 70/30 formula-level
in-distribution train/test split stratified within SATLIB configuration groups.
Thus, formulas from the same SATLIB configuration may appear in both training
and testing sets. Cross-family tests are reported separately to evaluate
generalization under distribution shift.

\begin{table}[tbh]
\setlength{\tabcolsep}{3.0pt}
\centering
\caption{Family-level summary of hybrid success.}
\label{tab:family_failure_summary}
\begin{tabular}{|l|r|r|r|r|}
\hline
\textbf{Family} & \textbf{Good} & \textbf{Conf.} & \textbf{Prop.} & \textbf{Rescue} \\ \hline
CBS & 85.6\% & 81.5\% & 79.6\% & 0.0\% \\ \hline
RTI & 82.0\% & 82.5\% & 79.9\% & 0.0\% \\ \hline
BMS & 63.0\% & 39.6\% & 39.8\% & 100.0\% \\ \hline
flat & 48.3\% & 22.8\% & 19.2\% & 0.0\% \\ \hline
sw & 15.7\% & 0.0\% & -1.8\% & 0.0\% \\ \hline
\end{tabular}
\end{table}

\subsection{SATLIB Family-Level Behavior}

Table \ref{tab:satlib_selected_improvements} summarizes the median conflict and
propagation counts for selected CBS clause-density buckets, together with RTI
and BMS results. Across the CBS families, pure CDCL requires median conflict
counts between 214 and 286.5, while the hybrid method reduces these medians to
between 32 and 55. This corresponds to conflict reductions of
$80.8\%$--$85.5\%$. Median propagation counts are reduced from
4943.5--6693.5 to 802.5--1289.5, corresponding to propagation reductions of
$80.2\%$--$84.6\%$.

The RTI family shows a similar trend, with an $83.4\%$ median conflict
reduction and an $83.1\%$ median propagation reduction. These results indicate
that p-bit-derived assumptions can substantially reduce CDCL search effort on
selected satisfiable random and controlled-backbone 3-SAT instances.

Figures \ref{fig:cbs_conflicts} and \ref{fig:cbs_props} show the CBS trends as
a function of clause density $\alpha=m/n$ and controlled backbone size $b$.
For the evaluated CBS configurations, the hybrid curve remains consistently
below the pure-CDCL curve for both conflicts and propagations. This suggests
that, within these selected CBS settings, p-bit-derived assumptions often guide
CDCL toward productive regions of the search space.

The BMS family shows more moderate aggregate improvement, with a $37.8\%$
median conflict reduction and a $41.3\%$ median propagation reduction. However,
the median rescue rate for BMS is 1.0. Therefore, at least half of the BMS
formulas route all five hybrid seeds to unrestricted rescue rather than being
solved during the assumption-guided attempts. This indicates that the BMS
improvement should not be attributed solely to direct p-bit assumption quality.
Instead, it is consistent with a warm-restart effect, where CDCL reuses learned
clauses generated during failed guided attempts. A separate attribution
baseline is needed to isolate this effect.

The graph-coloring families are less favorable to the proposed guidance
mechanism. The flat graph-coloring family has a $48.3\%$ hybrid-good rate and
a $22.8\%$ median conflict reduction. The sw family is a stronger failure case:
it has only a $15.7\%$ hybrid-good rate, $0.0\%$ median conflict reduction, and
$-1.8\%$ median propagation reduction. Thus, on sw instances, the hybrid method
slightly increases the median propagation count. Interestingly, sw instances
have the highest median agreement statistic, $q_{\mathrm{abs}}=0.638$, among
the evaluated families. This shows that strong p-bit sample polarization alone
does not guarantee useful CDCL assumptions.

\subsection{Classifier-Gated Hybrid Solving}

\begin{table}[tbh]
\setlength{\tabcolsep}{3.0pt}
\centering
\caption{Leakage-contaminated diagnostic gate summary.}
\label{tab:gate_summary}
\begin{tabular}{|l|r|r|r|r|r|}
\hline
\textbf{Policy} & \textbf{Apply} & \textbf{Keep} & \textbf{Avoid} & \textbf{Conf.} & \textbf{Prop.} \\ \hline
Always Hybrid & 100.0\% & 100.0\% & 0.0\% & 67.6\% & 54.8\% \\ \hline
Probe RF & 87.5\% & 94.8\% & 41.0\% & 67.4\% & 55.1\% \\ \hline
Two-Stage RF & 87.3\% & 94.6\% & 41.0\% & 67.2\% & 54.0\% \\ \hline
Always Pure & 0.0\% & 0.0\% & 100.0\% & 0.0\% & 0.0\% \\ \hline
\end{tabular}
\end{table}

Because p-bit guidance is highly distribution-sensitive, we also evaluate
learned suitability gates that decide whether a formula should be routed to the
hybrid solver or to pure CDCL. Table \ref{tab:gate_summary} summarizes the
gating results. The intended role of the gate is not necessarily to improve
aggregate performance on a hybrid-friendly benchmark mixture, but to preserve the
most beneficial hybrid applications while avoiding harmful ones, especially on
families such as sw.

The always-hybrid policy reduces total conflicts by $67.6\%$ and total
propagations by $54.8\%$ relative to always-pure CDCL. This reflects the fact
that the selected benchmark mixture is dominated by hybrid-friendly CBS
instances. The random-forest ProbeGate achieves $83.9\%$ accuracy, $86.3\%$
precision, and $94.8\%$ recall. It preserves $94.8\%$ of hybrid wins while
routing $87.5\%$ of formulas to the hybrid path. The gate avoids $41.0\%$ of
hybrid-loss cases.

However, these gating results should be interpreted as diagnostic rather than
deployment-ready. The feature set contains Oracle-derived and benchmark metadata
features, discussed below. Therefore, the reported gate performance is best
viewed as a leakage-contaminated upper bound on the learnability of hybrid
suitability under the current experimental protocol.

\section{Conclusion}
\label{sec:conc}

This paper studied a p-bit-guided CDCL framework that connects physics-inspired 
stochastic search with conventional SAT solving. The proposed method uses an Ising 
sampler to identify stable high-agreement literals, passes them to CDCL as temporary 
assumptions, and preserves correctness through unrestricted CDCL fallback. The 
exploratory ProbeGate results suggest that hybrid suitability may be learnable, 
but stricter evaluation is needed before deployment. Future work should remove 
feature-leakage sources, add attribution baselines, evaluate the method on broader 
SAT families and security-motivated CNFs, and extend the framework to quantified 
and constraint-rich domains.

\bibliographystyle{ieeetr}
\bibliography{references}

\end{document}